\documentclass[aps,prc,preprint]{revtex4}
\usepackage{graphicx}
\usepackage{bm}
\begin{document}
\setcounter{page}{1}\pagestyle{myheadings}
\begin{center}
{\bf Retardation of Particle Evaporation from Excited Nuclear Systems\\
 Due to Thermal Expansion\\}
\bigskip
\bigskip
{J. T\~oke, L. Pie\'nkowski, M. Houck, and W.U.
Schr\"oder\\}
{\it Department of Chemistry, University of Rochester,
Rochester, New York 14627\\}
\bigskip
{L.G. Sobotka\\} {\it Department of Chemistry, Washington
University, St. Louis, Missouri 63130\\}
\end{center}
\bigskip
\bigskip
\centerline{ABSTRACT} \begin{quotation}Particle evaporation rates
from excited nuclear systems at equilibrium matter density are
studied within the Harmonic-Interaction Fermi Gas Model (HIFGM)
combined with Weisskopf's detailed balance approach. It is found
that thermal expansion of a hot nucleus, as described
quantitatively by HIFGM, leads to a significant retardation of
particle emission, greatly extending the validity of Weisskopf's
approach. The decay of such highly excited nuclei is strongly
influenced by surface instabilities.
\end{quotation}

\newpage
\section{Introduction}

Studies of nuclear heavy-ion reactions at Fermi bombarding
energies (a few tens of MeV per nucleon) have faced conceptual and
experimental challenges. This is the domain in which the
low-energy dissipative reaction mechanism \cite{RedBook} is
expected to morph into one that is characterized by a merging of
time scales for collective motion and relaxation processes and the
population of a new phenomenological realm \cite{Bor90}. For
example, for nuclear temperatures of $T \sim 5 MeV$, nucleon
evaporation times of the order of $t_{evap}\approx 20 fm/c$ have
been estimated \cite{Bor90, Dan84}. Such short evaporation times
describe a ``prompt'' nuclear decay occurring during the nuclear
interaction between projectile and target, thus preventing a
complete equilibration of the system. Furthermore, mechanical and
chemical instabilities \cite{LiS01} expected for highly excited
nuclear systems should exceed the range of conventional
statistical models and cause novel nuclear decay modes
\cite{Gro90, Bon95}.

On the other hand, experimental observations
\cite{Lott,Djerroud,IWMF_03} of the particle emission patterns in
reactions at Fermi energies, induced by heavy projectiles,
demonstrate the persistence of sequential decay of hot
projectile-like fragments and their target-like reaction partners.
Apparently, the primary hot fragments emit nucleons, light
particles, and complex nuclear clusters with patterns
characteristic of thermalized sources. While the presence of an
additional, ``intermediate-velocity source'' component in the
particle emission patterns (most likely, of dynamical origin)
\cite{ourIVS, Montoya, Romualdo} tends to complicate the analysis,
the existence of intense sequential evaporation components
demonstrates unexpectedly high (meta-) stability of the primary
hot projectile and target-like nuclei.

The present work investigates conditions for a consistency of the
apparently long particle emission times with high nuclear
excitations. It is shown that, consistent with the above
observations, highly excited nuclear systems are indeed similar to
a classical compound nucleus, once their expansion to the
equilibrium matter density \cite{toke_surfentr} is taken into
account. The particle emission timescales are then long enough to
allow for a meaningful equilibration of the system in the sense of
Weisskopf's detailed balance \cite{Weiss37}. Relative decay rates
for various statistical decay channels are governed by the entropy
at a respective ``transition'' state, in which the state of the
surface plays an important role. The transition state may be
identified with an emission barrier for particle emission and a
separation saddle for statistical fission-like decay.

In the following section, the Weisskopf evaporation model is
briefly reviewed. The effects of nuclear expansion are described
in Section III, followed by a presentation of calculations in
Sect. IV and brief conclusions (Sect. V).

\section{Weisskopf's Particle Evaporation Time-scales}

The evaporation time scales for (metastable) excited nuclear
systems can be conveniently evaluated using Weisskopf's detailed
balance \cite{Weiss37} approach. This approach relates the
probability $W(E^*_A,\epsilon)$ for the emission of a particle
with energy $\epsilon$ by a compound nucleus with mass number $A$
and excitation energy $E^*_A$ to the cross section
$\sigma(\epsilon)$ for the reverse process of absorption of such a
particle by a nucleus of mass number $B$ and corresponding
excitation energy $E^*_B$,

\begin{equation}
W(E^*_A,\epsilon)d\epsilon=\sigma(\epsilon){gm\epsilon\over
\pi^2\hbar^3}{\omega_B(E^*_B)\over \omega_A(E^*_A)}d\epsilon,
\label{eq:balance}
\end{equation}

\noindent where $\omega_A$ and $\omega_B$ are the level densities of
nuclei $A$ and $B$ at respective excitation energies, $m$ is the
mass of the emitted particle, $\hbar$ is Planck's constant divided
by $2\pi$, and $g$ denotes the spin degeneracy.

In Eq.~\ref{eq:balance}, excitation energies $E^*_A$ and $E^*_B$ of
the compound nucleus $A$ and the residue $B$ are related via energy
conservation

\begin{equation}
E^*_B=E^*_A-\epsilon-Q_{gg},
\end{equation}

\noindent where $Q_{gg}$ is the $Q-value$ for the particle
separation from the ground state of nucleus $A$, leaving the residue
$B$ also in its ground state.

By expressing level densities $\omega$ in terms of entropy $S$

\begin{equation}
\omega_A(E^*_A)=e^{S_A(E^*_A)}\;\;\; and\;\;\;
\omega_B(E^*_B)=e^{S_B(E^*_B)}
\end{equation}

one rewrites Eq.~\ref{eq:balance} in a form

\begin{equation}
W(E^*_A,\epsilon)d\epsilon=\sigma(\epsilon){gm\epsilon\over
\pi^2\hbar^3}e^{S_B(E^*_A-Q_{gg}-\epsilon)-S_A(E^*_A)}d\epsilon,
\label{eq:balancemain}
\end{equation}
which is well suited for use in conjunction with models such as
HIFGM \cite{toke_surfentr,toke_box}. Ultimately, the decay width
$\Gamma$ for a particular particle channel is obtained by
integrating Eq.~\ref{eq:balancemain} over particle energy
$\epsilon$ and then by multiplying the result by $\hbar$, i.e.,

\begin{equation}
\Gamma(E^*_A)=\hbar\int_0^\infty W(E^*_A,\epsilon)d\epsilon={gm\over
\pi^2\hbar^2}\int_0^\infty \sigma(\epsilon)\epsilon
e^{S_B(E^*_A-Q_{gg}-\epsilon)-S_A(E^*_A)}d\epsilon. \label{eq:gamma}
\end{equation}

To be able to make use of Eq.~\ref{eq:gamma} one needs additionally
a prescription for calculating the cross section as a function of
bombarding energy. In the context of Weisskopf's approach, one
assumes customarily that the absorption cross section is equal to
the geometrical cross section of the nucleus. This corresponds to a
constant value of $\sigma=\pi r_o^2$ for neutrons and an
energy-dependent, Coulomb corrected expression for charged
particles,
\begin{equation}
\sigma=\pi r_o^2(1-{V_C\over\epsilon}), \label{eq:sigma}
\end{equation}

\noindent where $r_o$ is the nuclear radius parameter and $V_C$ is
the magnitude of the Coulomb barrier. Obviously, Eq.~\ref{eq:sigma}
is suited for both, charged particles and neutrons, with $V_C=0$ for
the latter.

A useful approximation for the right-hand side of
Eq.~\ref{eq:gamma} can be obtained \cite{Weiss37} for the neutron
decay width. It assumes an identical functional dependence of the
entropy on thermal excitation energy for both, mother and daughter
nucleus, i.e., $S_A(x)$=$S_B(x)$. Then, expanding the entropy
$S_B$ in a Taylor series about $E^*_A$ to first order results in

\begin{equation}
S_B(E^*_A-Q_{gg}-\epsilon) - S_A(E^*_A)=-{dS_A\over
dE^*_A}*(Q_{gg}+\epsilon)=-{1\over T_A}(Q_{gg}+\epsilon),
\label{eq:approxentropy}
\end{equation}

\noindent where use was made of the fact that the derivative of
the entropy with respect to excitation energy, $dS_A/dE^*_A$, is equal to the
inverse of the temperature, $T_A$, of nucleus A.
Thus, for the neutron decay width one obtains an approximate
expression

\begin{equation}
\Gamma_n(E^*_A)={gm\over \pi\hbar^2}r_o^2T_A^2e^{-{Q_{gg}\over
T_A}}, \label{eq:gammaapprox}
\end{equation}

\noindent revealing the strong temperature dependence of the decay
width. It is this exponential temperature dependence that has
raised concerns regarding the applicability of Weisskopf's
equilibrium approach at higher excitation energies. Rationale
given in Sect. III will demonstrate a much weaker effective
temperature dependence of the decay width. In figures shown
further below, the decay rate is expressed in terms of the decay
time scales parameter $\tau$, which is related to the decay width
$\Gamma$ via Heisenberg's uncertainty principle,
$\tau=\hbar/\Gamma$.

\section{Effects of Thermal Expansion of Nuclear Matter}

Like most liquids, nuclear matter is expected to respond to
thermal excitation by expanding, so as to arrive at a state of
maximum entropy for any given total excitation energy. While
thermal expansion should be predicted by any realistic model of
bound nuclear matter, a particularly simple expression for the
dependence of the asymptotic, equilibrium matter density on total
excitation energy is provided by the Harmonic Interaction Fermi
Gas Model (HIFGM). The essentials of this model
\cite{toke_surfentr,toke_box} are described below.

The central notion of the HIFGM is entropy. The relation between this
entropy and the thermal excitation energy $E^*_{therm}$ is assumed to be described by the
regular Fermi-gas expression, such that
\begin{equation}
S(E^*_{tot})=2\sqrt{a(E^*_{tot}-E^*_{compr})}, \label{eq:entropy}
\end{equation}

\noindent where $E^*_{tot}$ is the total excitation energy and
$E^*_{compr}$ is its collective compressional (potential energy) part.
The difference between these latter two energies defines the random, thermal
excitation $E^*_{therm}$ and, hence, the entropy. Furthermore, the
HIFGM adopts the matter density dependence of the level
density parameter $a$ (``little-a'') germane to the Fermi-gas model,
\begin{equation}
a=a_o({\rho\over\rho_o})^{-{2\over 3}}, \label{eq:littlea}
\end{equation}

\noindent where $a_o$ is the ground-state value of the
level-density parameter and $\rho_o$ is the ground-state matter
density. For simplicity, the HIFGM assumes a quadratic dependence
of the compressional energy, $E^*_{compr}$, on the relative matter
density \cite{eesm}
\begin{equation}
E^*_{compr}=E_{bind}(1-{\rho\over\rho_o})^2, \label{eq:compr}
\end{equation}

\noindent where $E_{bind}$ is the ground-state binding energy of the
system.

Given Eqs.~\ref{eq:entropy}-\ref{eq:littlea}, the equation for
maximum entropy as a function of relative density can be resolved
analytically, yielding \cite{toke_surfentr}

\begin{equation}
{\rho_{eq}\over \rho_o} = {1\over 4}(1+\sqrt{9-{E^*_{tot}\over
E_{bind}}}). \label{eq:eqdens}
\end{equation}

\noindent Assessing qualitatively the effects of thermal expansion
on the decay width (see Eq.~\ref{eq:gammaapprox}), one notes that,
according to Eq.~\ref{eq:eqdens}, the matter density decreases
with increasing total excitation energy. Consequently, the
compressional part of the excitation energy and the level density
parameter, both increase. \textit{The latter two trends reduce the
rate of increase of the system temperature with increasing
excitation energy, as compared to that without expansion.} This
follows from the Fermi-gas model relationship between temperature
($T$), level-density parameter ($a$), and thermal part
($E^*_{therm}$) of the excitation energy,

\begin{equation}
E^*_{therm}=E^*_{tot}-E^*_{compr}=aT^2. \label{eq:t}
\end{equation}
This effect can be viewed as a manifestation of the
$LeCh$$\hat{a}$$telier $ $Principle$. The reduction in effective
temperature from the value that it would have at normal matter
density leads to a reduction of the decay width, as approximated
by Eq.~\ref{eq:gammaapprox}. This reduction is partially offset by
the effects of $a)$ an increased value of the inverse capture
cross section ($\sigma$) and $b)$ a neutron binding energy
($Q_{gg}$) that is reduced by an amount equal to the compressional
energy per nucleon. Results of calculations performed using the
``exact'' Eq.~\ref{eq:gamma} are discussed in the next section.

\begin{figure}
\includegraphics{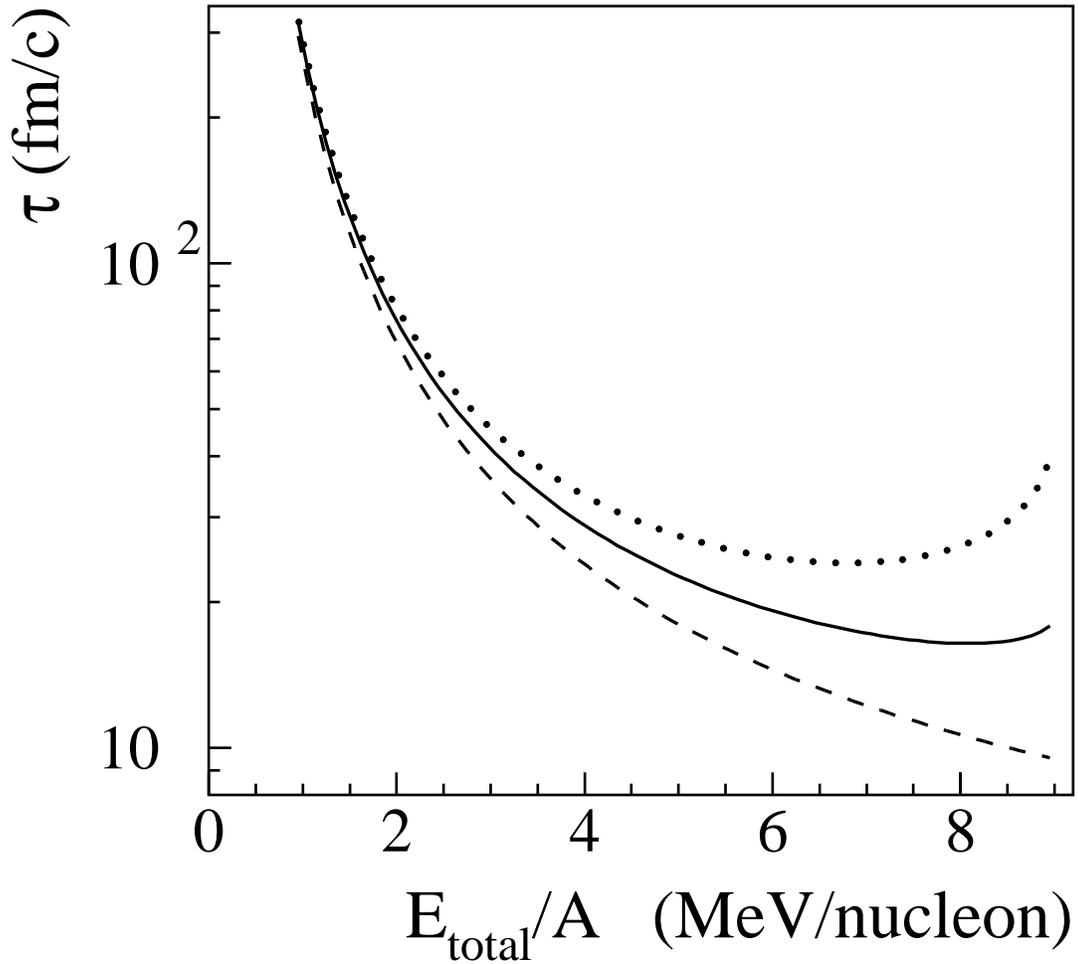}
\caption{Excitation-energy dependence of Weisskopf's nucleon
evaporation time scales for nuclei at equilibrium (solid line) and
ground-state (dashed line) matter density. The dotted line
illustrates results of calculations with $T=0$ absorption cross
section.}\label{fig:Retard}
\end{figure}

\section{Results of Calculations}

Results of model calculations for the Weisskopf particle
evaporation time scales are shown in Figs. 1 and 2. In Fig. 1, the
evolution of the average nucleon evaporation time $\tau$ with
total excitation energy per nucleon, predicted for expanded Fermi
matter at equilibrium density (solid line), is compared to that
resulting for Fermi matter forced to stay at ground-state density
(dashed line). As seen in this figure, if allowed to expand to
equilibrium density, a nucleus excited to $E^*_{tot}/A$$\geq 8$
$MeV$ regains a degree of (meta)stability against nucleon
evaporation comparable to that of a ground-state density nucleus
at the much lower excitation of $E^*_{tot}/A$$\sim 4.5$ $MeV$. The
small increase in $\tau$ predicted to occur near the
high-excitation end of the curve is related to the negative heat
capacity predicted by the HIFGM in that energy domain. Both these
effects are due to the fact that the considered maximum in entropy
is not absolute, but conditional, obtained by imposing a certain
functional form on the radial distribution of the nuclear matter
density.

\begin{figure}
\includegraphics{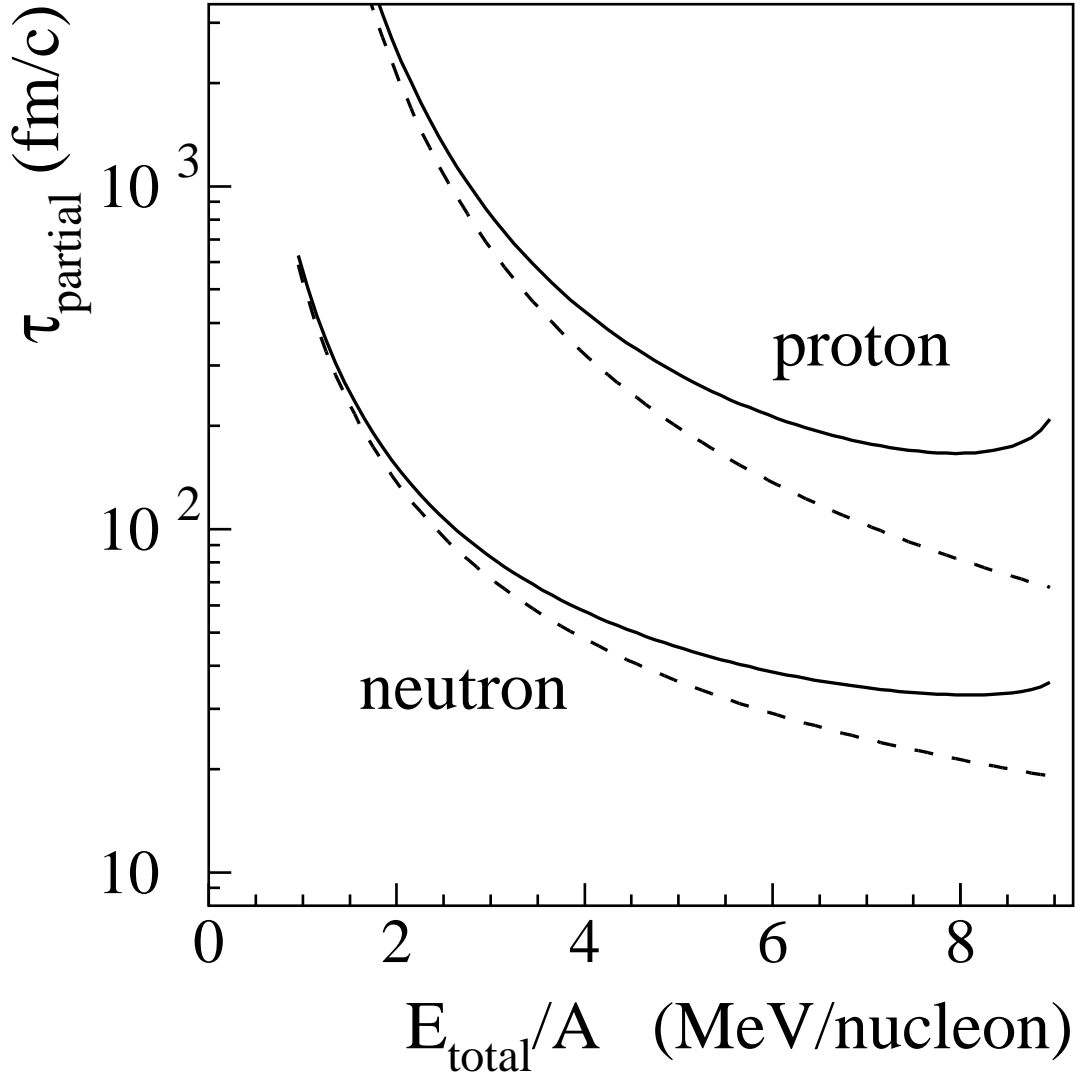}
\caption{Excitation-energy dependence of Weisskopf's neutron (bottom
curves) and proton (top curves) partial evaporation time scales for
nuclei at equilibrium (solid line) and ground-state (dashed line)
matter density.}\label{fig:Retard_partial}
\end{figure}

In Fig.~2, partial time scales are illustrated separately for
neutron emission and for proton emission. In both cases, thermal
expansion is seen to lead to a retardation of emission, extending
the range of applicability of Weisskopf's approach.

\section{Conclusions}
\label{sec:conclusions} The present calculations have shown that thermal
expansion of a nucleus to equilibrium density has a stabilizing
effect on the excited system, such that the most likely statistical
decay channels are suppressed. This may give the system time to
reach a more complete statistical equilibrium, where various surface modes
are excited as well. Such surface modes may be viewed as doorway states
associated with cluster emission. It has been demonstrated
elsewhere\cite{toke_surfentr} that, when equilibrated at
high excitation energies, such fission-like decay modes may compete
successfully with nucleon emission. It should also be kept in mind that,
with every neutron that escapes before these cluster emission modes are
equilibrated, the probability for the emission of subsequent neutrons
decreases, due to isospin effects, until it equals that
for proton emission. This isospin effect adds additional credibility
to the idea that Weisskopf's approach may be applicable at much
higher nuclear excitation energies than commonly thought.

It is worth noting that, in its present schematic form, the HIFGM
assumes that thermal expansion occurs in a self-similar fashion,
i.e., that this expansion can be reduced to a rescaling of the
radial coordinate. In addition, for the sake of simplicity, it is
assumed that the effective nucleonic mass is equal to the free
nucleon mass. A more complete modelling \cite{sob2004,sob2005}
calls for more thorough and detailed treatment of the diffuse
nuclear surface and its evolution with increasing excitation. In
such a treatment, it would be highly desirable to account for the
finite range of the effective nuclear interaction. This finite
range may alter the thermostatic properties of surface matter with
respect to those of bulk matter at equal density and temperature.
Furthermore, it is necessary to include in a comprehensive
modelling also thermal surface oscillations of the hot nuclear
system, as the role of such oscillations is expected to increases
\cite{toke_surfentr} with increasing excitation energy (through
decreasing surface tension or increasing surface entropy). It is
to be expected that, in line with $LeCh$$\hat{a}telier's $
$Principle$, both, the bulk nuclear matter and the surface region
will evolve in a concerted manner that increases entropy and
stabilizes the overall system with respect to single nucleon
emission even further.

\begin{acknowledgments}
This work was supported by the U.S. Department of Energy grant No.
DE-FG02-88ER40414.
\end{acknowledgments}

\end{document}